\documentclass[final,leqno]{siamltex}
\usepackage{epsfig}
\usepackage{amsmath}
\usepackage{amssymb}

\usepackage{tikz}
\usepackage{graphicx}
\usepackage{subcaption}
\usepackage{float}
\usepackage{booktabs}  

\usepackage[notcite,notref]{showkeys}
\usepackage{hyperref}

\newtheorem{remark}{Remark}

\def\3bar{{|\hspace{-.02in}|\hspace{-.02in}|}}

  \numberwithin{equation}{section}
\numberwithin{table}{section} \numberwithin{figure}{section}

\title{Tracking mob Dynamics in online social networks Using epidemiology model based on Mobility Equations}
\author{
	 Jumana H. S. Alkhalissi\thanks{College of Computer Science and Information Technology, University of Al-Qadisiyah, Al- Diwaniyah, 58001, Iraq(email:\href{mailto:Jumana Alkhalissi} {jumana.alkhalissi@qu.edu.iq, jumanaalsady@gmail.com})
	}
 \and Ahmed Al-Taweel\thanks{Mathematical Sciences, Georgia Southern University, Statesboro, GA 30458, USA (email:\href{mailto:Ahmed Al-Tawee}{aaltaweel@georgiasouthern.edu})
	}
    }

\begin{document}

\maketitle

\begin{abstract}
Nowadays, social media is the main tool in our new lives. The outbreak news and all related obtained from social media, and mob events affect the of spread these news fast. Recently, epidemiological models to study disease spread and analyze the behavior of mob groups by dealing with "contagions" that propagate through user networks.
In this research, we introduced a mathematical model to analyze social behavior related to COVID-19 spread by examining Twitter activity from April 2020 to June 2020. 
The main feature of this model is the integration of mobility dynamics that be derived from the above real data, to adjust the rate of outbreak based on the response of social interactions. Consider mobility as a parameter of time-varying, and fluctuations in the rate of contact that is driven by factors like personal behavior or external affecting such as "lockdown" and "quarantine" etc., to track public sentiment and engagement trends during the pandemic. The threshold number is derived, and the existence of bifurcation and the stability of the steady states are established. Numerical simulations and sensitivity analysis of relevant parameters are also carried out. 
\end{abstract}


\pagestyle{myheadings}

\section{Introduction}\label{Section:Introduction}
The event of mob propagation has been studied across numerous fields, including sociology, communication studies, and computational modeling. In \cite{11} explained how people in a crowd can behave differently than they would independently, with anonymity and emotional contagion playing key roles. Recently in \cite{22}, authors developed theories for the digital world, showing how these factors still impact behavior online.

Social media has increasingly become a double-edged sword in facilitating both constructive and destructive forms of group mobilization. On the one hand, platforms can bring people together for creative or community-oriented activities, but on the other, they can serve as a catalyst for harmful mob events. The UK Anti-Migrant Riots in August 2024 illustrates how misinformation spread through social media can lead to widespread chaos. False claims about a stabbing incident fueled far-right protests across the UK, escalating into violent anti-migrant demonstrations and multiple arrests. These incidents highlight the potent role of social media in amplifying mob behavior, emphasizing the critical need for robust content moderation and the responsible use of these platforms to mitigate their misuse.

During the COVID-19 pandemic, Twitter played a pivotal role in shaping public discourse and mobilizing collective behavior, both supporting and opposing public health measures. While many users leveraged the platform to share accurate health information and promote adherence to safety protocols such as social distancing and \textbf{lockdowns}, a significant wave of resistance emerged, fueled by misinformation, skepticism, and political motivations. Hashtags like \textbf{\#EndTheLockdown}, \textbf{\#ReopenAmerica}, and \textbf{\#FreedomOverFear} trended widely, organizing anti-lockdown protests and amplifying conspiracy theories. This mob behavior frequently translated into real-world actions, including protests that defied social distancing mandates and escalated into confrontational encounters. Furthermore, the spread of misinformation undermined public health messaging, complicating efforts to maintain compliance with safety measures. These dynamics highlighted Twitter’s dual role as both a disseminator of valuable information and a tool for mobilizing resistance, underscoring the critical need for digital literacy, content moderation, and robust public health communication strategies to mitigate the adverse impacts of social media during crises.

Understanding how information spreads on social media and its subsequent impact on public behavior is paramount. Epidemiological models, traditionally developed to study the spread of diseases, have been adapted to analyze social phenomena by treating ideas and sentiments as "contagions" propagating through user networks. These models allow researchers to capture the dynamics of user engagement, predict trends, and evaluate interventions. Epidemiological models, such as compartmental models, play a crucial role in organizing populations into distinct groups, providing a foundational mathematical framework for comprehending epidemic dynamics. \cite{33} introduced a fundamental SI model that classifies the population into susceptible and infected, extendable to an SIS model where individuals in the infected compartment can revert to a susceptible state \cite{44,55,66}. The SIR model, a widely recognized epidemic model \cite{77,88,99}, categorizes individuals into three compartments: susceptible, infected, and recovered. The SEIZ model (susceptible, exposed, infected, skeptic) has been employed to simulate the spread of news and rumors on the Twitter platform \cite{1010,1111,1212}. Mathematical modeling plays an important role in comprehending and providing useful methods to predict and control the dynamics of infectious diseases \cite{1313}. Numerous scholars have used the infectious disease dynamics model for smoking, alcoholism, drug addiction, game addiction, social media addiction, and other issues \cite{1414,1515,1616,1717,1818,1919,2020}. In \cite{2121} Anderson discuss mathematical models for the spread of infectious diseases,  and how mobility (or movement between compartments) plays a crucial role in disease transmission dynamics. 
In this work, we applied a SQCIR model to analyze social behavior related to COVID-19 by examining Twitter activity between April 2020 and June 2020. Our analysis focused on key terms such as ``lockdown" and ``social distancing" to track public sentiment and engagement trends during the pandemic. In addition, we also used the mobility equation to model the mobber decision-making time, i.e., the time it takes the mobber to decide whether to act, withdraw, act against, or perform a power exchange. The mobility equation model has individuals move between different compartments or regions and their movement impacts the spread of infections. In epidemiology, a mobility equation helps model how diseases spread through populations by describing the movement of individuals between different states, such as susceptible, exposed, infected, and recovered. These equations are used to predict how quickly a disease might spread and how interventions might affect its transmission. To work with an epidemiology model, we typically define variables for each group of people and create equations to describe the rates at which individuals move between these groups. By solving these equations, you can simulate the disease's spread and understand how changes (like vaccination or social distancing) can reduce the number of infections over time. 

 The paper is organized as follows. Section \ref{secM} presents our methodology, including Twitter data and a mathematical framework for the epidemiological model, and focuses on the various components contributing to our understanding of the spread of mob event over time. Section \ref{sec3} is devoted to investigating model analysis including the basic reproduction number, the stability analysis, bifurcation, and sensitivity analysis of the model. The results from numerical analysis are reported in Section \ref{sec4} to validate the theoretical aspect of the model. Finally, concluding remarks and future research plans are given in Section \ref{cc}.

\section{Methodology}\label{secM}
This section provides an overview of data collection and the methodology used in this article.
\subsection{Data Collection}
Social media networks (Twitter, Telegram, Facebook, Instagram, Facebook, etc.) have developed into sites of intensive and constant information dissemination between government organizations, specialists, and the general public due to the COVID-19 epidemic global spread.  Numerous scientific research studies have indicated that social media platforms and other news websites may be valuable data sources for problem analysis and understanding individuals’ behavior during a pandemic over time \cite{2222,2323}. Numerous monitoring strategies have been performed to assist public health management and specialists in making decisions to identify enormous volumes of data from social media platforms. This data can be used to detect the ideas, attitudes, emotions, sentiments, and topics on which individuals' senses are concentrated in reaction to the COVID-19 virus \cite{2424}. Systematic analysis of the data aids policymakers and health professionals in identifying problems of public concern and resolving them most effectively. We collected approximately 2,35,240 tweets about the COVID-19 discourse on Twitter between April 19 and June 20, 2020. To get tweets collected, we use a developer account on Twitter that allows us to access Twitter API v2 to search tweets on specific topics. With the access option for the academic research project, we were able to search the archive database of tweets on Twitter based on keywords and date specifications. We use code samples from Twitter Developer Relations and the search-tweets-python package to aid us in tweet collections (https://github.com/twitterdev/search-tweets-python). All tweets were collected utilizing a set of COVID-19 related keywords, including ${``COVID-19''}$,  ${``coronavirus''}$, ${``corona''}$, ${``quarantine''}$, ${``homequarantine''}$, ${``quarantinecenter''}$, ${``socialdistancing''}$, ${``staysafe''}$, ${``covid''}$, ${``covaccine''}$, ${``lockdown''}$, ${``stayhome''}$.  The keywords were chosen based on their popularity on Google Trends during the COVID-19 life cycle of data collection. The parameters for the tweet search include data fields such as ${``tweet_{-}ID''}$,  ${``user_{-}mention''}$, ${``like_{-}count,''}$, ${``quote_{-}count''}$, ${``creation_data''}$, ${``favorite_{-}count''}$, ${``reply_{-}count''}$, ${``retweet_{-}count''}$, ${``source''}$, ${``tweet''}$, ${``hashtags''}$, ${``stayhome''}$. We pre-processed these collected data by setting a user-defined pre-processing function based on NLTK (Natural Language Toolkit, a Python library for NLP). At the initial stage, it converts all the tweets into lowercase. Then, it removes all extra white spaces, numbers, special characters, ASCII characters, URLs, punctuations and stopwords from the tweets. Then, it converts all `covid' words into `covid19' as we already removed all numbers from the tweets. The pre-processing function reduces the number of inflected words to their word stem, using stemming. We used NLTK's Sentiment Analyzer to calculate sentiment polarity for each cleaned and pre-processed tweet, generating scores for positive, negative, and neutral categories, along with a compound sentiment score. Tweets were then classified into Positive, Negative, or Neutral categories based on the compound score, and sentiment polarity ratings were assigned accordingly.
\subsection{The SQCIR Mob propagation model}

\begin{figure}[ht!]
	\centering
	\includegraphics[width=0.6\textwidth]{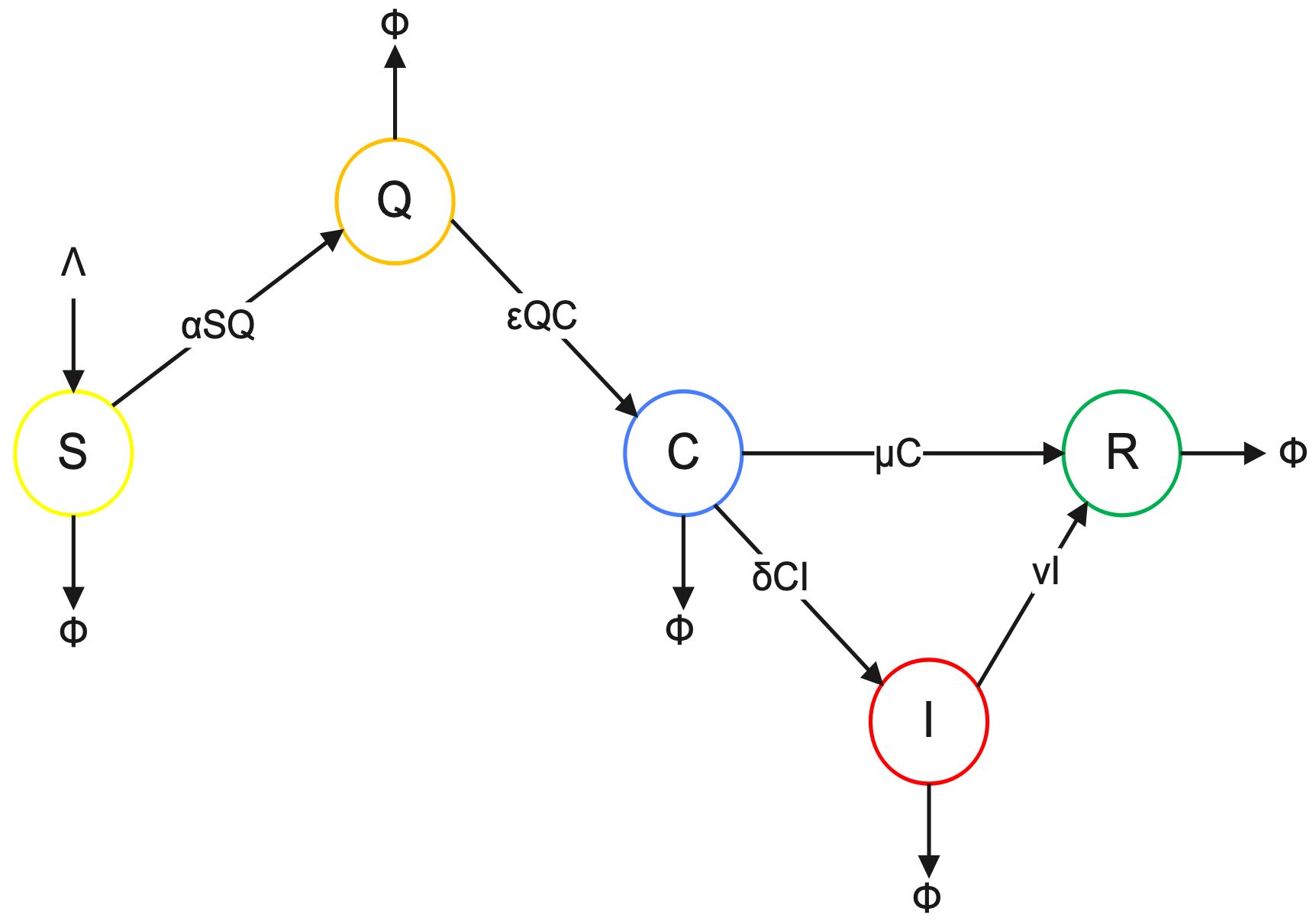} 
	\caption{SQCIR compartment model of mob propagation.}
	\label{SEIREq}
\end{figure}

A mob is an event managed through social media networks, email, SMS, or other digital communication technologies in which a group of people gathers online or offline to collectively conduct an act and then spread (fast or over time). Analyzing such a phenomenon is challenging due to a lack of data, theoretical methods, and resources. For this purpose, we employ epidemiological techniques to model the behavior of mobbers, utilizing the mobility equation for tracking mobber. The total population is divided into five compartments: $S_{i}(t)$, the number of susceptible individuals who have not encountered the mob; $Q_{i}(t)$, the number of individuals who are initially suspicious or influenced by media, temporally not spreading the mob; $C_{i}(t)$, the Individual who have encountered the mob but are not actively spreading it; $I_{i}(t)$, the Individual who believes and actively spread the mob; $R_{i}(t)$, the Individual who no longer spread the mob, either become they lost interest or encountered reliable information.
 Our model is controlled by the following system of five ordinary differential equations. A transfer diagram of the model is shown in Fig. \ref{SEIREq}, and the model's parameters are given in Table \ref{tab:seiz-parameters}.

In general, mobility can be defined as a network of
interacting communities where the connections and the interrelated intensity represent the movement among them \cite{2525}.
The model is constructed with the Mobility equation and Monte Carlo simulation  as follows:
\begin{equation}\label{eq111}
 	\begin{cases}
	\begin{split}
		\frac{dS_{i}}{dt}&= \Lambda-\alpha S_{i} Q_{i} -\epsilon(t) S_{i} C_{i}-\Phi S_{i}+\sum_{j}T_{ji}S_{j}-\sum_{j}T_{ij}S_{i},\\
		\frac{dQ_{i}}{dt}&=\alpha S_{i} Q_{i} -\epsilon(t) Q_{i}C_{i}-\Phi Q_{i}+\sum_{j}T_{ji}Q_{j}-\sum_{j}T_{ij}Q_{i},\\
		\frac{dC_{i}}{dt}&= \epsilon(t) S_{i}C_{i}+\epsilon(t) Q_{i}C_{i}-\delta C_{i} I_{i}-\mu C_{i}-\Phi C_{i}+\sum_{j}T_{ji}C_{j}-\sum_{j}T_{ij}C_{i},\\
		\frac{dI_{i}}{dt}&= \delta C_{i} I_{i}-v I_{i}-\Phi I_{i}+\sum_{j} T_{ji} I_{j}-\sum_{j} T_{ij} I_{i},\\
		\frac{dR_{i}}{dt}&= \mu  C_{i} +v I_{i}-\Phi R_{i}+\sum_{j} T_{ji} R_{j}-\sum_{j} T_{ij} R_{i},
	\end{split}
  	\end{cases}
\end{equation}
where $\epsilon(t)$ is the time varying contact rate that can be adjusted by mob dynamics through Monte Carlo simulations, the mobility term $\sum_{j} T_{ji} S_{j}$ represents people entering region $i$, while  $\sum_{j} T_{ij} S_{j}$ represents people leaving region $i$ and $\epsilon(t)=\epsilon(0)\cot (1- M_{t})-M_{t} $ is random factor between $0$ and $1.5.$ 
{ {\begin{table}[h]
			\centering
			\caption{Definitions of Parameter  OF the SQCIR model \ref{eq111}.}
			\label{tab:seiz-parameters}
				   \resizebox{1.\textwidth}{!}{ 
                   \begin{tabular}{ccc}
					\toprule
					Parameter & Definition& Estimate\\
					\midrule
					$\Lambda$ & The entering number of susceptible individuals per unit of time&4.00\\
     $\alpha$ & The rate at which susceptible individuals became quarantined due to suspicion or media effect&0.14 \\
      $\epsilon$&The rate at which quarantined individuals became contacted&0.03\\
      $\delta$ & The rate at which contacted individuals became infected&0.10\\
      $\mu$ & The rate at which contacted individuals recover without becoming infected&0.10\\
					$\nu$ & The rate at which infected individuals became recover&0.05\\
					
                    $\Phi$ &  The  emigration rate of moving out of the system after the population moves&0.01 \\
                   
					\bottomrule
			\end{tabular}}
\end{table}}}
For the sake of simplicity, we will confine our attention to the initial stages of a mob event in social media. The above system can be reduced as follows:
\begin{equation}\label{eq1}
 	\begin{cases}
	\begin{split}
		\frac{dS}{dt}&= \Lambda-\alpha S Q -\epsilon S C-\Phi S,\\
		\frac{dQ}{dt}&=\alpha S Q -\epsilon QC-\Phi Q,\\
		\frac{dC}{dt}&= \epsilon SC+\epsilon QC-\delta C I-\mu C-\Phi C,\\
		\frac{dI}{dt}&= \delta C I-v I-\Phi I,\\
		\frac{dR}{dt}&= \mu  C+v I-\Phi R.
	\end{split}
  	\end{cases}
\end{equation}

The total number of individuals on social networks $N(t)$ satisfies the following equation:
$$N(t) =S(t) +Q(t) +C(t) + I(t) + R(t),$$
with initial conditions $ S(0) \geq 0,  Q(0) \geq 0 ,  C(0) \geq 0 ,  I(0) \geq 0 , \text{ and }  R(0) \geq 0 $ at $ t = 0 .$
Then, we have
	\begin{equation}
		\frac{dN}{dt}\leq \frac{\Lambda}{\Phi}.\label{eq11}
	\end{equation}
	The solution of equation \ref{eq11} given by $N(t)\leq N(0)\exp(-\Phi t)+\frac{\Lambda}{\Phi}(1-exp(-\Phi t))$. Then, as $t\longrightarrow\infty, N(t)\longrightarrow\frac{\Lambda}{\Phi}$. Therefore, the model positively invariant region is given by:
		\begin{equation}
	\varSigma=\left\{\left(S(t),Q(t),C(t),I(t),R(t))\in \Re^{5}_{+}: 0< N(t)\leq \frac{\Lambda}{\Phi}\right)\right\}.
\end{equation}
\section{Model analysis}\label{sec3}
Now, we are able to present the following equilibrium points and basic reproduction number for Equations \ref{eq1}.

\subsection{Equilibrium points}
To get the equilibria, we seek the time-independent solutions $(S^{\prime}, Q^{\prime}, C^{\prime}, I^{\prime}, R^{\prime})$  that satisfy system \ref{eq1} with the time derivatives set to zero. The resulting system is as follows:
\begin{equation}\label{eeeq1}
 	\begin{cases}
	\begin{split}
		0&= \Lambda-\alpha S Q -\epsilon S C-\Phi S,\\
		0&=\alpha S Q -\epsilon QC-\Phi Q,\\
		0&= \epsilon SC+\epsilon QC-\delta C I-\mu C-\Phi C,\\
		0&= \delta C I-v I-\Phi I,\\
		0&= \mu  C+v I-\Phi R.
	\end{split}
  	\end{cases}
\end{equation}
By solving the system equations \ref{eeeq1}, in the absence of the mob of social media, we assume that $Q=C =I= 0$. Therefore, the mob free  equilibria (MFE)  is given by:
\begin{equation}\label{MFE}
	MFE=\left(S_{0},0,0,0,R_{0}\right) \text{ where }  S_{0}=\frac{\Lambda}{\Phi}.
\end{equation}
Next, the threshold  is computed by utilizing the next-generation matrix method.
\subsection{Basic reproduction number $\Re_{0}$}
In this section, we will use the basic reproduction number $\Re_0$ of the system \ref{eq1} to analyze the nature of mob propagation. We use the next-generation matrix method to calculate the basic reproduction number $\Re_0$ when the entire population is susceptible and has no contact and infection compartment. We will consider two different types of equations:
\begin{equation}
	\begin{cases}
		\frac{dC}{dt}&= \epsilon SC+\epsilon QC-\delta C I-(\mu +\Phi) C,\\
	\frac{dI}{dt}&= \delta C I-(v +\Phi) I.
	\end{cases}
	\label{eq:w2}
\end{equation}
From the above equation, we obtained
\begin{equation}\label{22}
F=	\begin{pmatrix}
		 \epsilon SC+\epsilon QC\\
		0
	\end{pmatrix},
\end{equation}
and 
\begin{equation}\label{12}
V=	\begin{pmatrix}
		\delta C I+(\mu+\Phi) C\\\\
	(v+\Phi) I- \delta C I
	\end{pmatrix}.
\end{equation}

The Jacobian matrices at MFE of the matrices in equations \ref{22} and \ref{12} is given as:

\begin{equation}\label{23}
F=	\begin{pmatrix}
	\frac{\Lambda\epsilon}{\Phi}&0 \\
		0&0
	\end{pmatrix},
\end{equation}
\begin{equation}\label{24}
V=	\begin{pmatrix}
		\Phi+\mu&0\\
		0&\Phi+v
	\end{pmatrix},
\end{equation}
\begin{equation}
	F\cdot  V^{-1}=	\begin{pmatrix}
	\frac{\Lambda\epsilon}{\Phi(\Phi+v)}&0\\
		0&0
	\end{pmatrix}.
\end{equation}
Hence, the threshold parameter $\Re_{0}$ is obtained by taking the largest eigenvalue 
\begin{equation}
	\Re_{0}=\frac{\Lambda\epsilon}{\Phi(\Phi+v)}.
\end{equation}
\begin{remark}
 If $\Re_{0}>1$, the infection is expected to spread in the population. The infection will die out if $\Re_{0}<1$. The basic reproduction number $\Re_{0}$, where accounting for mob events, the transmission rate is modified by a time-varying factor $M(t)$ leading to a time-varying reproduction number $\Re_{t}$
\begin{eqnarray}
    \epsilon(t)=\epsilon_{0}\cdot\left(1+M(t)\right).
\end{eqnarray}
\end{remark}
\subsection{Local Stability Analysis of Mob Free Equilibrium}
The analysis of the model’s equilibrium stability shows the following results.
\begin{theorem}
The MFE point is locally asymptotically stable if $\Re_{0}<1$ and unstable otherwise.
\end{theorem}
\proof We examine the Jacobian of model \ref{eq1}, given by:
	\begin{equation}\label{ja}
	J=	\begin{pmatrix}
		-C\epsilon-\Phi-Q\alpha& -S \alpha &-S\epsilon &0&0\\
		Q\alpha&-C\epsilon-\Phi +S \alpha &-Q\epsilon&0&0\\
		C\epsilon&C\epsilon&-I \delta -\Phi+Q\epsilon+S\epsilon-\mu&-C\delta&0\\
		0&0&I\delta&C\delta-\Phi-v&0\\
		0&0& \mu &v&-\Phi
	\end{pmatrix}.
\end{equation}
Evaluating equation (\ref{ja}) at the MEF point (\ref{MFE}), we obtain
	\begin{equation}
		J=	\begin{pmatrix}
		-\Phi& -\frac{\Lambda\alpha}{\Phi} &-\frac{\Lambda\epsilon}{\Phi} &0&0\\
		0&\frac{\Lambda\alpha}{\Phi}-\Phi  &0&0&0\\
		0&0&\frac{\Lambda\epsilon}{\Phi}-\Phi-\mu&0&0\\
		0&0&0&-\Phi-v&0\\
		0&0& \mu &v&-\Phi
	\end{pmatrix}.
	\end{equation}
	From the above Jacobian matrix, the negative eigenvalues are
	$\lambda_{1}=\lambda_{2} =-\Phi,  \lambda_{3}=\frac{\Lambda\alpha-\Phi^{2}}{\Phi} ,\lambda_{4}=\frac{\Lambda\epsilon-\Phi^{2}-\Phi\mu}{\Phi}$, and $\lambda_{5}= -\Phi-v$.
 Thus, the mob free equilibrium is asymptotically stable locally.
	
	\subsection{Existence of Endemic Equilibrium State}
The endemic equilibrium point of the model \ref{eq11} is identified when $ S\neq Q\neq C\neq I\neq R\neq 0$,
 denoted by $E^{\star} = (S^{\star}, Q^{\star}, C^{\star}, I^{\star}, R^{\star}) \neq 0$ and can be obtained by setting each equation of the system to zero, i.e.,
	\begin{equation}
		\frac{dN}{dt}=	\frac{dS}{dt}=	\frac{dQ}{dt}=	\frac{dC}{dt}=	\frac{dI}{dt}=	\frac{dR}{dt}=0.
	\end{equation}
	Then, we obtain
	\begin{equation}\label{EPE}
 	\begin{cases}
S^{\star}&=\frac{\epsilon(v+\Phi)+\delta\Phi}{\alpha\delta},\\
	Q^{\star}&=\frac{\Lambda\delta}{\epsilon(v+\Phi)+\delta\Phi}-\frac{\epsilon(v+\Phi)}{\alpha\delta}-\frac{\Phi}{\alpha},\\
	C^{\star}&=\frac{v+\Phi}{\delta},\\
	I^{\star}&=\frac{\epsilon}{\delta}\left(\frac{\epsilon(v+\Phi)+\delta\Phi}{\alpha\delta}+\frac{\Lambda\delta}{\epsilon(v+\Phi)+\delta\Phi}-\frac{\epsilon(v+\Phi)}{\alpha\delta}-\frac{\Phi}{\alpha}\right)+\frac{\mu+\Phi}{\delta},\\
	R^{\star}&=1-\left[(1+\frac{\epsilon}{\delta})\left(\frac{\epsilon(v+\Phi)+\delta\Phi}{\alpha\delta}+\frac{\Lambda\delta}{\epsilon(v+\Phi)+\delta\Phi}-\frac{\epsilon(v+\Phi)}{\alpha\delta}-\frac{\Phi}{\alpha}\right)+\frac{(v+\mu+2\Phi)}{\delta}\right].
 	\end{cases}
\end{equation}	
The system equation (\ref{EPE}) demonstrates that if $\Re_{0}>1$, then the endemic equilibrium  $ E^{\star} = (S^{\star}, Q^{\star}, C^{\star}, I^{\star}, R^{\star})\in \Sigma.$
\subsection{Bifurcation of endemic equilibrium point}
\begin{theorem}
    If $\Re_{0}<1$ the mob free equilibrium $MFE_{0}$ of system \ref{eeeq1} is locally asymptotically stable and it is unstable if $\Re_{0}>1$.
\end{theorem}
\proof The Jacobian matrix of contacted and infected compartments is 
\begin{equation}
		J_{MFE_{0}}=	\begin{pmatrix}
		\frac{\Lambda \epsilon}{\Phi}&0\\
		0&-(\nu+\Phi)
	\end{pmatrix}.
	\end{equation}
The determinant of $J_{MFE_{0}}$ obtained from characteristic equation of the matrix as 
\begin{eqnarray}\label{123}
    \det(J_{MFE_{0}})=-\left(\frac{\Lambda\epsilon}{\Phi}-(M+\phi)\right)(\Phi+v).
\end{eqnarray}
According to the assumption that $\Re_{0}<1$, we hae 
\begin{eqnarray}
    \frac{\Lambda}{\Phi(\Phi+\nu)}<1  \rightarrow \Lambda<\Phi(\Phi+\nu).
\end{eqnarray}
Then,
\begin{eqnarray}
    (M+\Phi)(\epsilon\Phi +\nu)>\frac{\Lambda\epsilon}{\Phi}(\Phi+\nu).
\end{eqnarray}
Conversely if $\Re_{0}>1$, we get 
\begin{eqnarray}
    (M+\Phi)(\epsilon\Phi +\nu)>\frac{\Lambda\epsilon}{\Phi}(\Phi+\nu) \textit{due to } \Lambda>\Phi (\Phi+\nu),
\end{eqnarray}
Thus according to Routh- Hurwitz criteria, all possible solution of equation (\ref{123}) have negative real part iff $\Re_{0}<1$, as a result of mob free equilibrium point is locally asymptotically stable.

To identify the threshold for spread infections or die out, we need to know the critical relationship between affected parameters, as: let $\Re_{0}=1,$ then
\begin{eqnarray}
    \frac{\epsilon\Lambda}{\Phi(\Phi+\nu)}=1\Longrightarrow \epsilon_{c}=\frac{\Phi(\Phi+\nu)}{\Lambda},
\end{eqnarray}
if $\epsilon>\epsilon_{c}$, the infection persists, and if $\epsilon <\epsilon_{c}$ the mob- free equilibrium is stable and about critical rate $\Lambda_{c}$:
\begin{eqnarray}
    \Lambda_{c}=\frac{\Phi(\Phi+\nu)}{\epsilon}.
\end{eqnarray}
This tells us the minimum number of susceptible individuals need to be infection continuous at given $\epsilon$. Moreover, critical rate of $\Phi$:
\begin{eqnarray}
    \Phi_{c}=\frac{\epsilon\Lambda}{\nu+\Phi},
\end{eqnarray}
a higher emigration rate makes it harder for the infection to continuous.
\subsection{Sensitivity analysis}
\begin{figure}[H]
	\centering
	\includegraphics[width=0.6\textwidth]{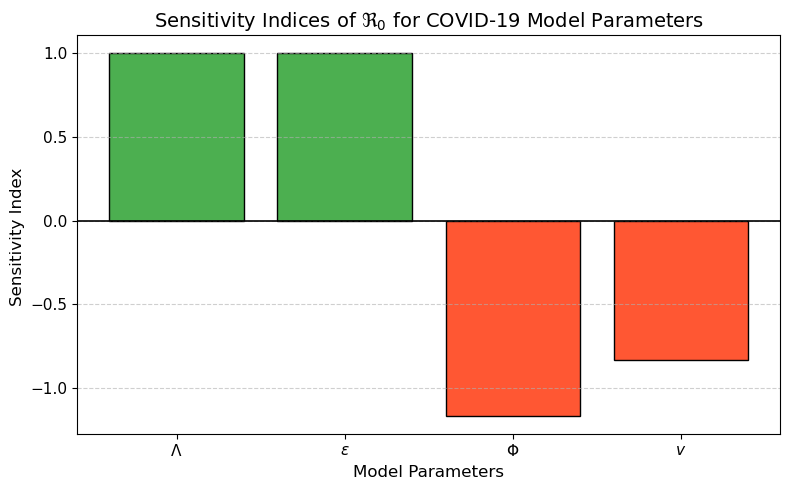} 
	\caption{The figure shows the sensitivity indices of $\Re_0$ for the dependent parameters of the COVID19 model \ref{eq1}.}
	\label{SEIjjjgggggmmREq}
\end{figure}

We performed a sensitivity analysis to show the effect of each parameter on the mob transmission. The sensitivity indices concerning the parameter values are given in the form of

\begin{eqnarray}\label{eqf}
	\Pi_{\rho}^{\Re_{0}}=\frac{d\Re_{0}}{d\rho}\times\frac{\rho}{\Re_{0}},
\end{eqnarray}
where $\rho$ represents all the basic parameters and 
\begin{equation}
	\Re_{0}=\frac{\Lambda\epsilon}{\Phi(\Phi+v)},
\end{equation}
\begin{eqnarray}
	\Pi_{\Lambda}^{\Re_{0}}=\frac{d\Re_{0}}{d\Lambda}\times\frac{\Lambda}{\Re_{0}}=1>0,
\end{eqnarray}
\begin{eqnarray}
	\Pi_{\epsilon}^{\Re_{0}}=\frac{d\Re_{0}}{d\epsilon}\times\frac{\epsilon}{\Re_{0}}=1>0,
\end{eqnarray}
\begin{eqnarray}
	\Pi_{\Phi}^{\Re_{0}}=\frac{d\Re_{0}}{d\Phi}\times\frac{\Phi}{\Re_{0}}=-\frac{2\Phi + v}{\Phi + v}<0,
\end{eqnarray}
\begin{eqnarray}\label{eql}
	\Pi_{v}^{\Re_{0}}=\frac{d\Re_{0}}{dv}\times\frac{v}{\Re_{0}}=-\frac{v}{\Phi + v}<0.
\end{eqnarray}
The numerical values showing the relative importance of $\Re_0$ parameters are shown in equations (\ref{eqf})-(\ref{eql}). As one can see, $\Lambda$ and $ \epsilon$ clearly affect the mob stability. Therefore, by increasing $\Lambda$ and $ \epsilon$ by 1\%, $\Re_0$ would increase by 1\% in influence. Some parameters have a positive relation, while others have a negative relation.
A negative relation suggests that increasing the metric's values would help reduce mob event brutality. At the same time, a positive relationship indicates that an increase would greatly influence the frequency of the mob event in the values of those parameters. Figure \ref{SEIjjjgggggmmREq} shows the sensitivity indices for the dependent parameters of basic reproduction number.
\begin{figure}[H]
	\centering
	\includegraphics[width=0.7\textwidth]{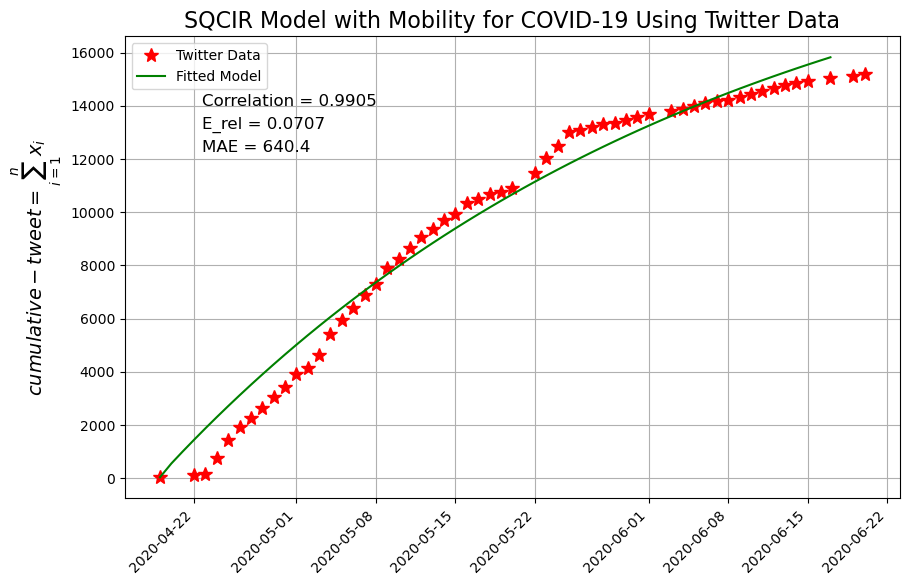} 
	\caption{Best fit modeling for mob Dynamics in social media networks. The figure shows the estimated infected case data fitting with the real infected case data of the SQCIR model \ref{eq111}.}
	\label{Sen}
\end{figure}

\section{Numerical simulations and discussions}\label{sec4}
This section proposes some numerical tests to validate the theoretical results. We apply the SQCIR model to find the numerical solution in the computation. The numerical experiment uses a Twitter dataset related to the COVID-19 epidemic from April 2020 to June 2020. The dataset comprised approximately 2,35,240 tweets interacting with the COVID-19 virus during the life cycle. To analyze individual transitions in different compartments of the SQCIR model for the Twitter dataset, we applied a least squared method  as given by:
	{{\begin{equation*}
		\Theta = \text{argmin}_{\varphi}\left(\sum_{i=0}^{\text{n}}\left(I_{cu}(t_{i})-\underbrace{I_{vcu}(t_{i})}_\text{tweet}\right)^{2}\right),
	\end{equation*}}}
	here, $\Theta$ denotes the vector containing estimated parameter values, $\varphi$ represents the parameter space, $t_{\text{n}}$ is the most recent date considered in the analysis, $t_{i}$ denotes the date, $I_{cu}(t_{i})$ signifies the cumulative incidence up to $t_{i}$, and $I_{vcu}(t_i)$ represents the cumulative incidence according to the model up to $t_{i}$. In order to evaluate the SQCIR models, Figures \ref{Sen} show how well the SQCIR models \ref{eq111} fit the COVID-19 on the Twitter data, along with the respective 2-norm relative error
	$E_{- \text{rel}} = \frac{\left\|I_{cu}(t_{i}) - I_{vcu}(t_{i})\right\|_{2}}{\left\|I_{vcu}(t_{i})\right\|_{2}}$, and the mean absolute error (MAE) described by: $MAE = \frac{\left\|I_{cu}(t_{i}) - I_{vcu}(t_{i})\right\|_{1}}{n}.$	The numerical examples indicate that the SQCIR model \ref{eq111} accurately fits Twitter data for the COVID-19 virus. The low error in the SQCIR model suggests a precise representation of Twitter data. The SQCIR model consistently captures the initial spread more effectively on Twitter, a phenomenon attributed to a delay as individuals in the exposed category take time before sharing their posts.
	\subsection{Analyzing the Reproduction Number $\Re_0$}
\begin{figure}[H]
    \centering
    \begin{subfigure}{0.45\textwidth}
        \centering
        \includegraphics[width=\textwidth]{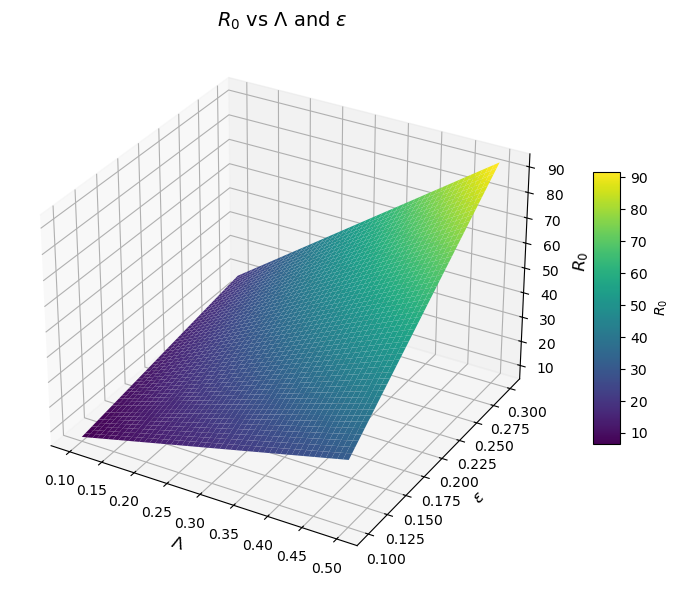}
        \label{fig:surface}
    \end{subfigure}
    \hfill
    \begin{subfigure}{0.45\textwidth}
        \centering
        \includegraphics[width=\textwidth]{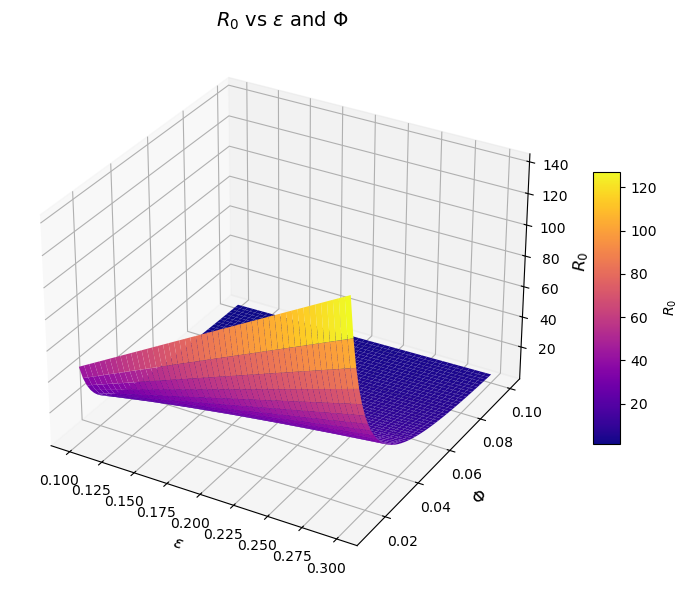}
        \label{fig:contour1}
    \end{subfigure}
    \vspace{0.5cm}
    \begin{subfigure}{0.45\textwidth}
        \centering
        \includegraphics[width=\textwidth]{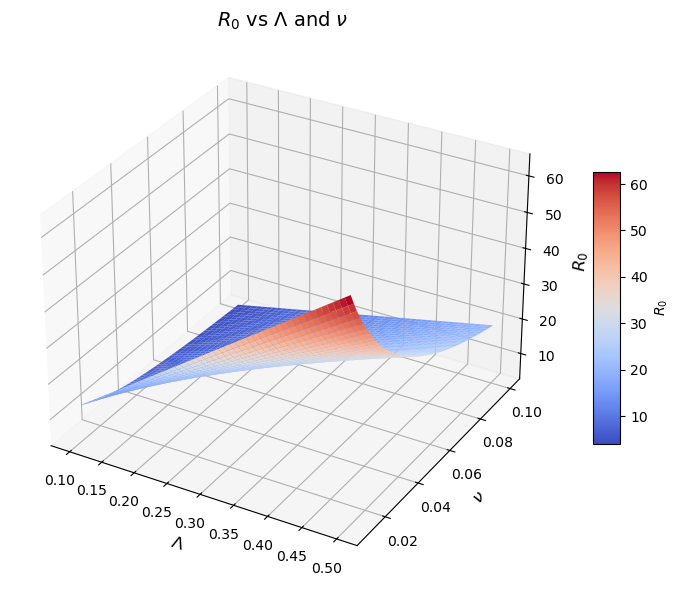}
        \label{fig:contour2}
    \end{subfigure}
    \hfill
    \begin{subfigure}{0.45\textwidth}
        \centering
        \includegraphics[width=\textwidth]{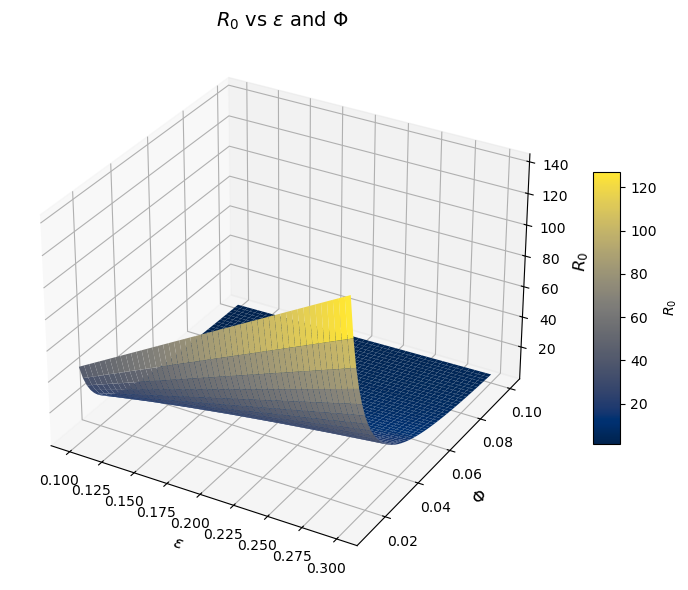}
        \label{fig:contour3}
    \end{subfigure}
    \caption{Dynamics of $\Re_0$ on different parameters.}
    \label{SEIjjjREq}
\end{figure}

This test uses relevant contour plots to explain the impact of critical parameters in the SQCIR model \ref{eq111} on the reproductive number $\Re_0$. By plotting the trajectories of sentiment spread across all the classes or compartments on social media platforms over time, we can gain a comprehensive understanding of the progression of sentiment posts. Figure \ref{SEIjjjREq} shows the numerical simulation supports the analytical results and highlights the relevance of the reproduction number $\Re_0$. These simulations also enhance our understanding of how different people interact on the social media platform. The top most sensitive parameters affecting $\Re_0$ are $\Lambda$, $\Phi$, $\epsilon$, and $\nu$. Figure \ref{SEIjjjREq} shows that $\Re_0$ decreases with a decrease in $\Lambda$ and $\nu$ and increases with other parameters.   
\subsection{Simulation of The SQCIR Model}
\begin{figure}[H]
    \centering
    \begin{subfigure}{0.6\textwidth}
        \centering
        \includegraphics[width=\textwidth]{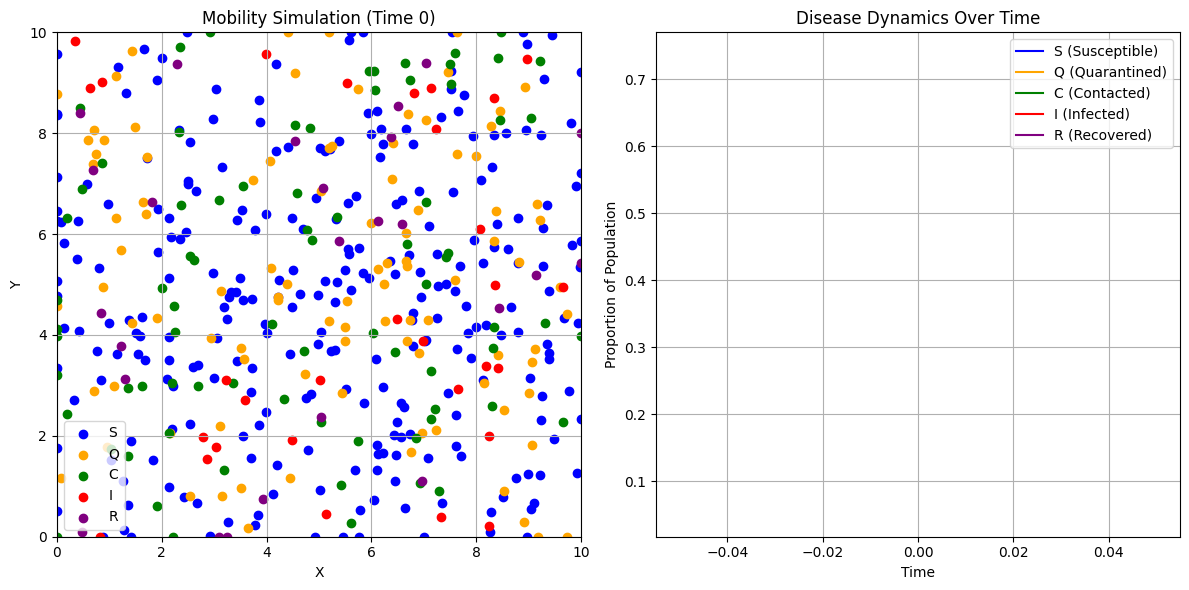}
        \label{fig:simulation}
    \end{subfigure}
    \hfill
    \begin{subfigure}{0.6\textwidth}
        \centering
        \includegraphics[width=\textwidth]{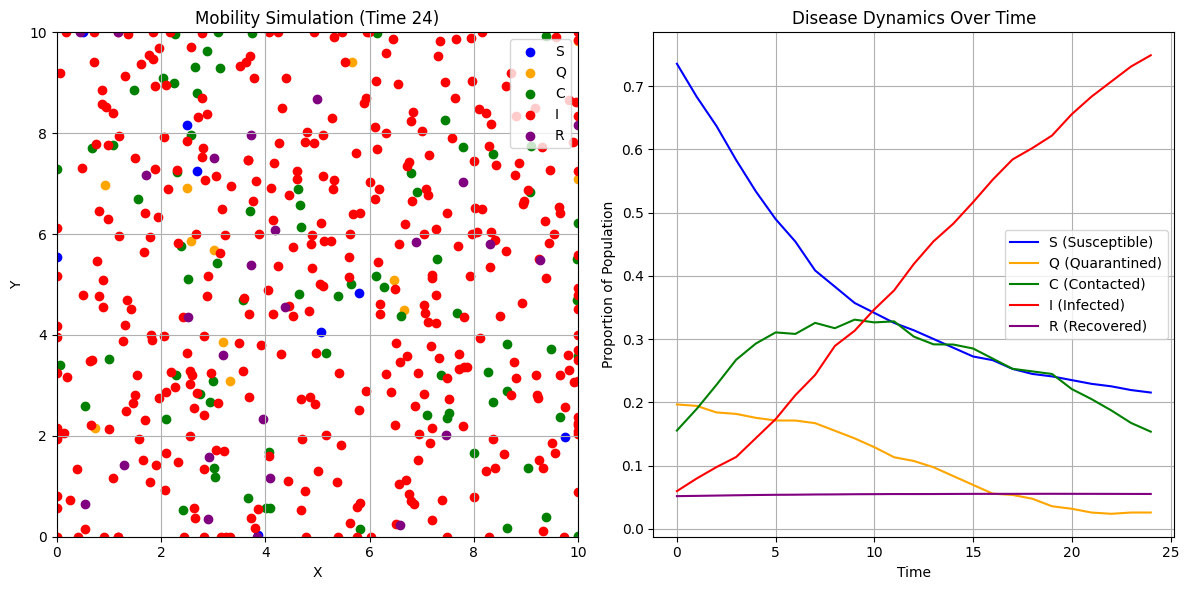}
        \label{fig:sim1}
    \end{subfigure}
    \vspace{0.cm}
    \begin{subfigure}{0.6\textwidth}
        \centering
        \includegraphics[width=\textwidth]{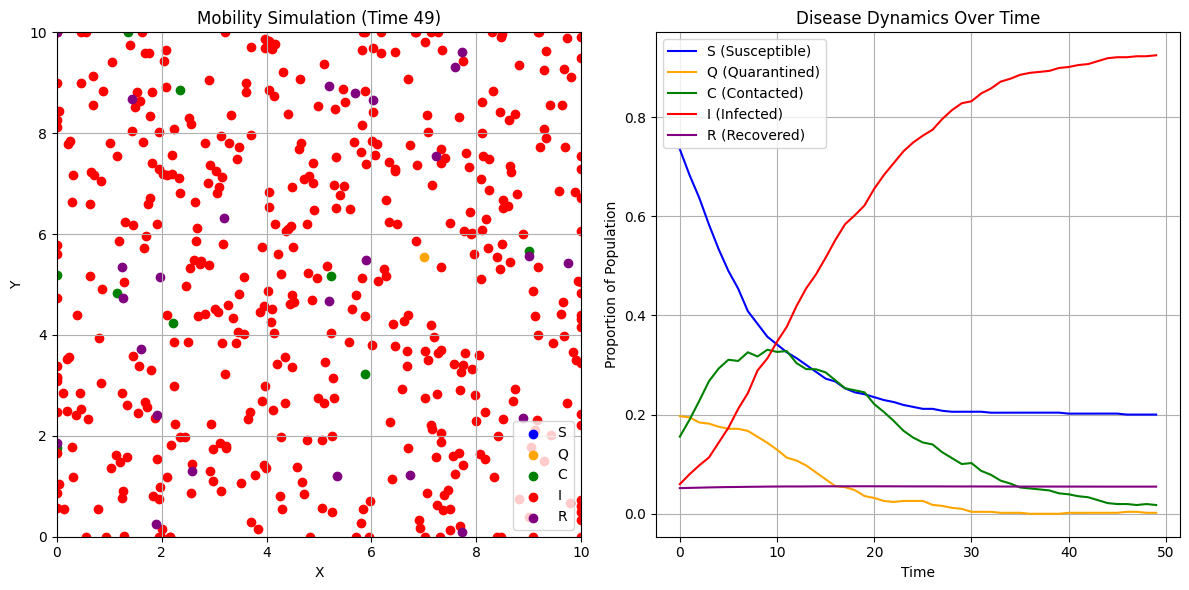}
        \label{fig:sim2}
    \end{subfigure}
    \hfill
    \begin{subfigure}{0.6\textwidth}
        \centering
        \includegraphics[width=\textwidth]{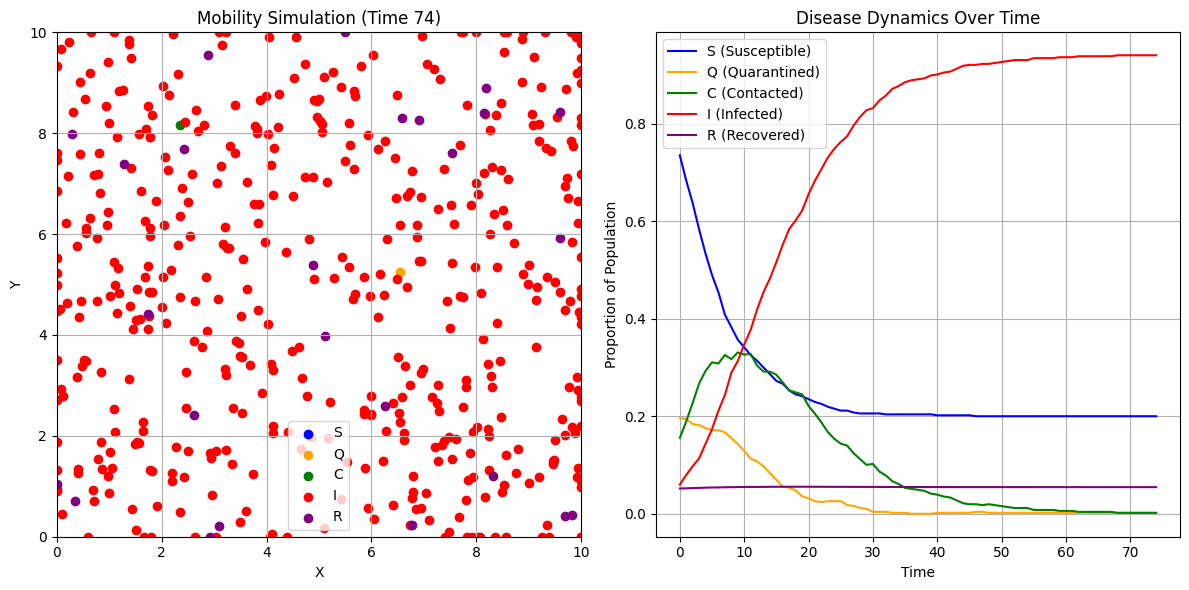}
        \label{fig:sim3}
    \end{subfigure}
    \begin{subfigure}{0.6\textwidth}
        \centering
        \includegraphics[width=\textwidth]{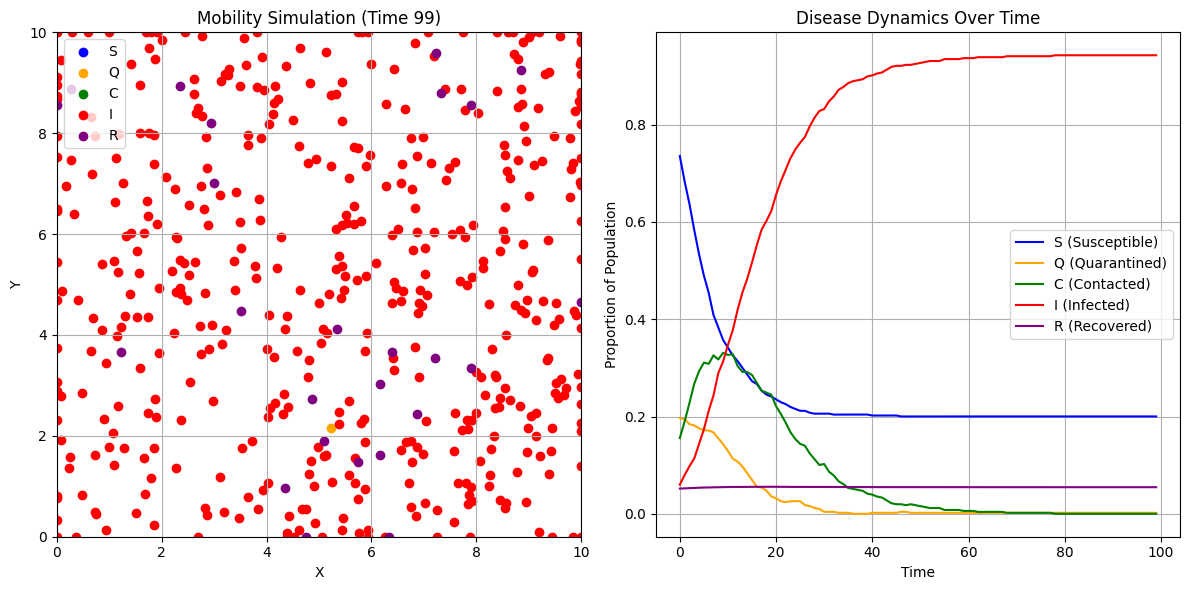}
        \label{fig:sim4}
    \end{subfigure}
    \caption{Simulation of the proposed model \ref{eq111} with mobility and disease dynamics over the time.}
    \label{SEM}
\end{figure}

In order to examine the infected by the proposed model, we discuss the numerical simulations. We consider simulation results from different iterations and dynamical behavior in the SQCIR model \ref{eq111}. As results depicted in \ref{SEM} confirm that the numbers of infected individuals grew over time, when quarantined and contacted numbers slowed down. Intuitively, contact tracing in the network and isolate the infectious nodes. Then it will reduce the rate of outbreaks as much as possible. 

Fig.\ref{SEM} shows the effect of mobility simulation in the population and what is the change in the behavior of spread outbreaks with parameters value ($\Lambda =2, \alpha =0.14, \epsilon= 0.26, \Phi =0.0074, \delta= 0.10, \mu =0.10, v = 0.05)$.  

\begin{figure}[H]
	\centering
	\includegraphics[width=0.9\textwidth]{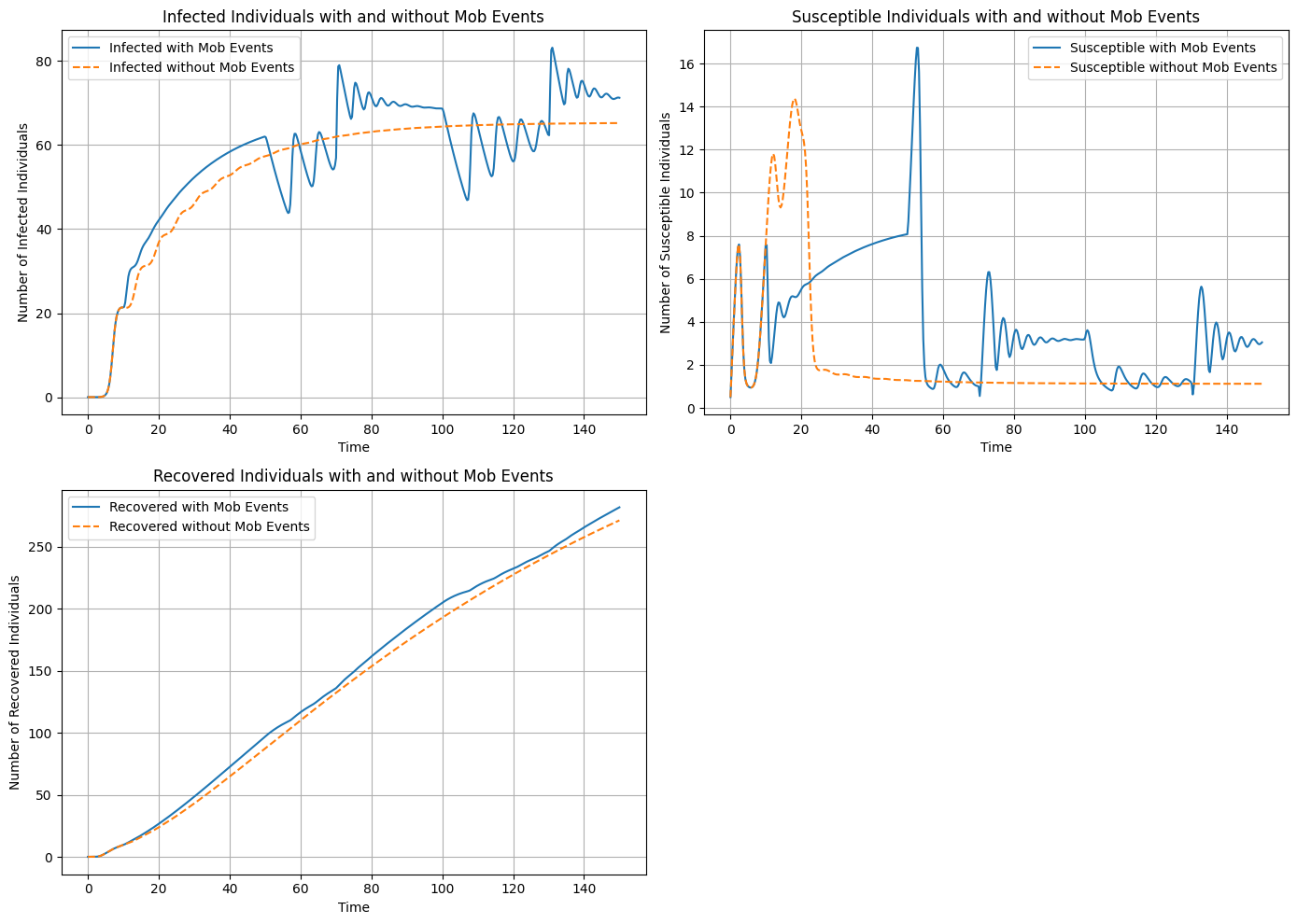} 
	\caption{Mob events and the affecting in the population with $\Large =4, \alpha= 0.14, \epsilon = 0.26, \Phi= 0.0074, \delta=0.10, \mu =0.10, v =0.05.$ }
	\label{peak}
\end{figure}

\begin{figure}[H]
	\centering
	\includegraphics[width=0.9\textwidth]{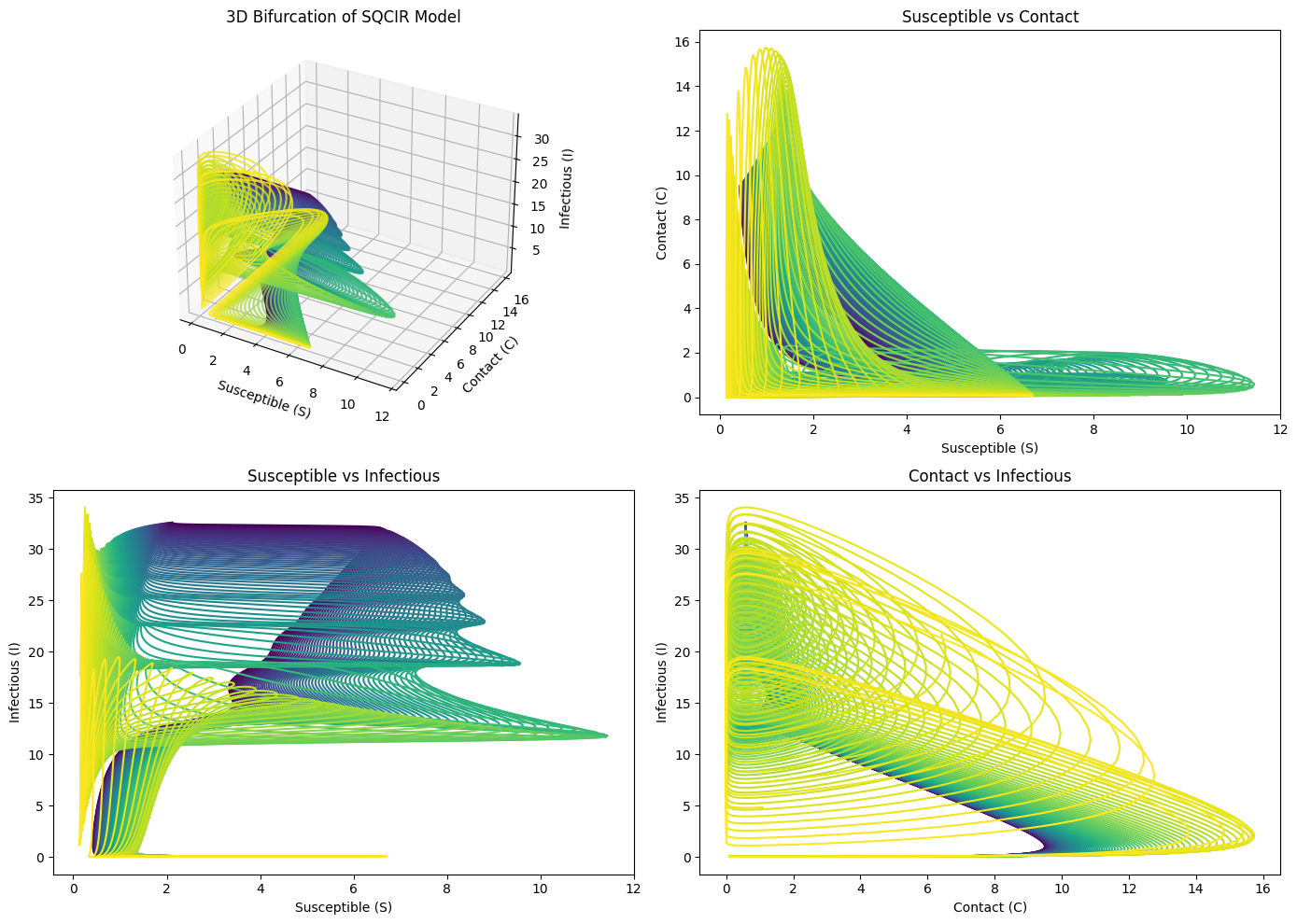} 
	\caption{Bifurcation of SQCIR model and the affect of different values of $\epsilon$. }
	\label{bifur}
\end{figure}

Fig.\ref{peak}, further investigates the effectiveness of the proposed model to explain the spread of the disease with mob event and without mob event. The results show that existing mob events increased the rate of infection and recover while the rate decreased if there were no events of this type as shown in Table \ref{tab:epidemic_metrics} :\\
\begin{table}[H]
\centering
\caption{Comparison of Epidemic Metrics With and Without Mob Events}
\label{tab:epidemic_metrics}
\begin{tabular}{|l|c|c|}
\hline
\textbf{Metric}                  & \textbf{With Mob Events} & \textbf{Without Mob Events} \\ \hline
\textbf{Peak Infected}           & 83.15 at time 131.06     & 65.21 at time 150.0         \\ \hline
\textbf{Duration of Epidemic}    & 144.59                   & 144.59                      \\ \hline
\textbf{Average Recovery Rate}   & 1.88                     & 1.81                        \\ \hline
\textbf{Total Infections}        & 28376.31                 & 27205.78                    \\ \hline
\end{tabular}
\end{table}
To shed light, the bifurcation analysis of the proposed model and the affect of the parameter $\epsilon$, we consider a small experimental result here. In Fig.\ref{bifur}, an increase of the parameter $\epsilon$ means that more individuals return to increasing in the contacted state, then increase the rate of infectious and raises the basic reproduction number $R_0$. However, the bifurcation movement to the right means strong transmission to sustain the spread of infection. As a result, $\epsilon$ demonstrates how the spread of the disease and the alteration between growing and being controlled.
  
\section{Conclusions}\label{cc}
In this paper, we developed a mathematical model to describe the dynamics of social media mob events. We simulated approximately 2,35,240 tweets interacting with the COVID-19 virus during the life cycle using the epidemiology model based on the mobility equation. The model was validated by comparing the simulated mobs’ results to the ground-truth data we collected from Twitter.  Our analysis reveals that the disease-free equilibrium of the model is locally asymptotically stable when the associated reproduction number $\Re_0$ is less than one, and it becomes unstable otherwise. We computed the basic reproduction number $\Re_0$ and examined the stability of the equilibrium points in detail and the bifurcation of the endemic equilibrium point. The main goal is to build a theoretical model that can help us better understand the mob phenomenon and aid policymakers in combating misinformation by targeting interventions toward skeptical users and understanding mob events' impacts on public opinion and decision-making processes. As the future work plan, we shall extend the SQCIR model to other data and techniques including more control interventions to tackle mob spread on social media platforms. Furthermore, we suggest including neutral sentiments in the model to acquire more information on Twitter's contagious dynamics and explore the different mob suppression methods in online social media networks.
\section*{Compliance with Ethical Standards}
Conflict of interest No conflict of interest exists in the submission of this manuscript.


\begin{thebibliography}{99}

\bibitem{11}
L. GUSTAVE, Crowd: A study of the popular mind. FORGOTTEN BOOKS, 2018. 

\bibitem{22} G. Le Bon, The crowd a study of the popular mind. BEYOND BOOKS HUB, 2023.

\bibitem{33} K. A. Kabir, K. Kuga, and J. Tanimoto, “Analysis of sir epidemic model with information spreading of awareness,” Chaos, Solitons \& Fractals, vol. 119, pp. 118–125, 2019.

\bibitem{44} T. Liu, P. Li, Y. Chen, and J. Zhang, “Community size effects on epidemic spreading in
multiplex social networks,” PloS one, vol. 11, no. 3, p. e0152021, 2016.

\bibitem{55} W. Luo and W. P. Tay, “Finding an infection source under the sis model,” in 2013 IEEE international conference on acoustics, speech and signal processing, pp. 2930–2934, IEEE,
2013.

\bibitem{66} H. Shi, Z. Duan, and G. Chen, “An sis model with infective medium on complex networks,”
Physica A: Statistical Mechanics and its Applications, vol. 387, no. 8-9, pp. 2133–2144,
2008.

 \bibitem{77}M. M. Nika, “Synthedemic modelling and prediction of internet-based phenomena,” 2014.

 \bibitem{88}N. A. Kudryashov, M. A. Chmykhov, and M. Vigdorowitsch, “Analytical features of the
sir model and their applications to covid-19,” Applied Mathematical Modelling, vol. 90,
pp. 466–473, 2021.

\bibitem{99} C. Ji and D. Jiang, “Threshold behaviour of a stochastic sir model,” Applied Mathematical
Modelling, vol. 38, no. 21-22, pp. 5067–5079, 2014.

\bibitem{1010} F. Jin, E. Dougherty, P. Saraf, Y. Cao, and N. Ramakrishnan, “Epidemiological modeling of
news and rumors on twitter,” in Proceedings of the 7th workshop on social network mining
and analysis, pp. 1–9, 2013.

 \bibitem{1111} A. Mathur and C. P. Gupta, “Dynamic seiz in online social networks: epidemiological mod-
eling of untrue information,” International Journal of Advanced Computer Science and
Applications, vol. 11, no. 7, 2020.

 \bibitem{1212}F. Jin, E. Dougherty, P. Saraf, Y. Cao, and N. Ramakrishnan, “Epidemiological modeling of
news and rumors on twitter,” in Proceedings of the 7th workshop on social network mining
and analysis, pp. 1–9, 2013.

 \bibitem{1313}J. D. Murray, Mathematical biology: I. An introduction, vol. 17. Springer Science \& Business Media, 2007.

 \bibitem{1414}H. T. Alemneh and N. Y. Alemu, “Mathematical modeling with optimal control analysis of
social media addiction,” Infectious Disease Modelling, vol. 6, pp. 405–419, 2021.

 \bibitem{1515}H.-F. Huo and X.-M. Zhang, “Modeling the influence of twitter in reducing and increasing the
spread of influenza epidemics,” SpringerPlus, vol. 5, pp. 1–20, 2016.

 \bibitem{1616}S. Yang, S. Liu, K. Su, and J. Chen, “A rumor propagation model considering media effect and suspicion mechanism under public emergencies,” Mathematics, vol. 12, no. 12, p. 1906,
2024.

 \bibitem{1717}H. Guo and X. Yan, “Dynamic modeling and simulation of rumor propagation based on the
double refutation mechanism,” Information Sciences, vol. 630, pp. 385–402, 2023.

 \bibitem{1818}Y. Tian and X. Ding, “Rumor spreading model with considering debunking behavior in emer-
gencies,” Applied Mathematics and Computation, vol. 363, p. 124599, 2019.

\bibitem{1919}L. Zhu and B. Wang, “Stability analysis of a sair rumor spreading model with control strategies
in online social networks,” Information Sciences, vol. 526, pp. 1–19, 2020.

 \bibitem{2020}F. Brauer, C. Castillo-Chavez, Z. Feng, et al., Mathematical models in epidemiology, vol. 32.
Springer, 2019.

  \bibitem{2121}R. M. May, Infectious diseases of humans: dynamics and control. Oxford University Press,
1991.

 \bibitem{2222}W. H. Organization et al., “Who director-general’s opening remarks at the media briefing on
covid-19,” January, vol. 30, 2022.

 \bibitem{2323}J. Csse, “Covid-19 data repository by the center for systems science and engineering (csse) at
johns hopkins university,” CSSE, Editor, 2020.

 \bibitem{2424}S. Liu and J. Liu, “Public attitudes toward covid-19 vaccines on english-language twitter: A
sentiment analysis,” Vaccine, vol. 39, no. 39, pp. 5499–5505, 2021.

 \bibitem{2525}D. Balcan, V. Colizza, B. Goncalves, H. Hu, J. J. Ramasco, and A. Vespignani, “Multiscale mobility networks and the spatial spreading of infectious diseases,” Proceedings of the
National Academy of Sciences, vol. 106, no. 51, pp. 21484–21489, 2009.

  \end{thebibliography}
\end{document}